\input harvmac

\def\K3{{\bf K3}}
\def\journal#1&#2(#3){\unskip, \sl #1\ \bf #2 \rm(19#3) }
\def\andjournal#1&#2(#3){\sl #1~\bf #2 \rm (19#3) }

\def\bar{\overline}

\def\tilde{\widetilde}

\def\frac#1#2{{#1\over#2}}

\def\inbar{\,\vrule height1.5ex width.4pt depth0pt}
\def\IC{\relax\hbox{$\inbar\kern-.3em{\rm C}$}}
\def\IR{\relax{\rm I\kern-.18em R}}
\def\IP{\relax{\rm I\kern-.18em P}}

%
%

%
\catcode`\@=11
\def\slash#1{\mathord{\mathpalette\c@ncel{#1}}}
\overfullrule=0pt

\def\underrel#1\over#2{\mathrel{\mathop{\kern\z@#1}\limits_{#2}}}

\catcode`\@=12


%

\def \sinh{{\rm sinh}}
\def \cosh{{\rm cosh}}


\input amssym
\input epsf

\lref\fzz{V.~Fateev, A.~B.~Zamolodchikov and A.~B.~Zamolodchikov,
``Boundary Liouville field theory. I: Boundary state and boundary
two-point function,'' arXiv:hep-th/0001012.
}

\lref\Teschner{ J.~Teschner, ``Remarks on Liouville theory with
boundary,'' arXiv:hep-th/0009138.
}

\lref\PolchinskiMY{ J.~Polchinski and L.~Thorlacius, ``Free
fermion representation of a boundary conformal field theory,''
Phys.\ Rev.\ D {\bf 50}, 622 (1994) [arXiv:hep-th/9404008].
}

\lref\TakayanagiGE{ T.~Takayanagi, ``Comments on 2D type IIA
string and matrix model,'' JHEP {\bf 0411}, 030 (2004)
[arXiv:hep-th/0408086].
}

\lref\KutasovPV{D.~Kutasov, ``Some properties of (non)critical
strings,'' arXiv:hep-th/9110041.
}

\lref\igor{ I.~R.~Klebanov, ``String theory in two-dimensions,''
arXiv:hep-th/9108019.
}

\lref\SeibergNM{ N.~Seiberg and D.~Shih, ``Branes, rings and
matrix models in minimal (super)string theory,'' JHEP {\bf 0402},
021 (2004) [arXiv:hep-th/0312170].
}

\lref\SeibergEB{ N.~Seiberg, ``Notes On Quantum Liouville Theory
And Quantum Gravity,'' Prog.\ Theor.\ Phys.\ Suppl.\  {\bf 102},
319 (1990).
}

\lref\joe{J.~Polchinski, ``What is String Theory?''
arXiv:hep-th/9411028
}

\lref\ginsparg{ P.~Ginsparg and G.~W.~Moore, ``Lectures On 2-D
Gravity And 2-D String Theory,'' arXiv:hep-th/9304011.
}

\lref\FukudaBV{ T.~Fukuda and K.~Hosomichi, ``Super Liouville
theory with boundary,'' Nucl.\ Phys.\ B {\bf 635}, 215 (2002)
[arXiv:hep-th/0202032].
}

\lref\korean{ C.~Ahn, C.~Rim and M.~Stanishkov, ``Exact one-point
function of $N = 1$ super-Liouville theory with boundary,'' Nucl.\
Phys.\ B {\bf 636}, 497 (2002) [arXiv:hep-th/0202043].
}

\lref\WittenZD{ E.~Witten, ``Ground ring of two-dimensional string
theory,'' Nucl.\ Phys.\ B {\bf 373}, 187 (1992)
[arXiv:hep-th/9108004].
}

\lref\CallanUB{ C.~G.~.~Callan, I.~R.~Klebanov, A.~W.~W.~Ludwig
and J.~M.~Maldacena, ``Exact solution of a boundary conformal
field theory,'' Nucl.\ Phys.\ B {\bf 422}, 417 (1994)
[arXiv:hep-th/9402113].
}

\lref\KlebanovWG{ I.~R.~Klebanov, J.~Maldacena and N.~Seiberg,
``Unitary and complex matrix models as 1-d type 0 strings,''
Commun.\ Math.\ Phys.\  {\bf 252}, 275 (2004)
[arXiv:hep-th/0309168].
}

\lref\SeibergEI{ N.~Seiberg and D.~Shih, ``Flux vacua and branes
of the minimal superstring,'' JHEP {\bf 0501}, 055 (2005)
[arXiv:hep-th/0412315].
}

\lref\BouwknegtYG{ P.~Bouwknegt, J.~G.~McCarthy and K.~Pilch,
``BRST analysis of physical states for 2-D gravity coupled to $c
\le 1$ matter,'' Commun.\ Math.\ Phys.\  {\bf 145}, 541 (1992);
}
\lref\ImbimboIA{
C.~Imbimbo, S.~Mahapatra and S.~Mukhi, ``Construction of physical
states of nontrivial ghost number in $c < 1$ string theory,''
Nucl.\ Phys.\ B {\bf 375}, 399 (1992).
}

\lref\BouwknegtAM{ P.~Bouwknegt, J.~G.~McCarthy and K.~Pilch,
``Ground ring for the 2-D NSR string,'' Nucl.\ Phys.\ B {\bf 377},
541 (1992) [arXiv:hep-th/9112036].
}

\lref\Itoh{ K.~Itoh and N.~Ohta, ``BRST cohomology and physical
states in 2-D supergravity coupled to c <= 1 matter,'' Nucl.\
Phys.\ B {\bf 377}, 113 (1992) [arXiv:hep-th/9110013];
``Spectrum of two-dimensional (super)gravity,'' Prog.\ Theor.\
Phys.\ Suppl.\  {\bf 110}, 97 (1992) [arXiv:hep-th/9201034].
}

\lref\BouwknegtVA{ P.~Bouwknegt, J.~G.~McCarthy and K.~Pilch,
``BRST analysis of physical states for 2-D (super)gravity coupled
to (super)conformal matter,'' arXiv:hep-th/9110031.
}

\lref\KutasovQX{ D.~Kutasov, E.~J.~Martinec and N.~Seiberg,
``Ground rings and their modules in 2-D gravity with $c\le 1$
matter,'' Phys.\ Lett.\ B {\bf 276}, 437 (1992)
[arXiv:hep-th/9111048].
}

\lref\BershadskyUB{ M.~Bershadsky and D.~Kutasov, ``Scattering of
open and closed strings in (1+1)-dimensions,'' Nucl.\ Phys.\ B
{\bf 382}, 213 (1992) [arXiv:hep-th/9204049].
}

\lref\KapustinHI{ A.~Kapustin, ``Noncritical superstrings in a
Ramond-Ramond background,'' arXiv:hep-th/0308119.
}

\lref\DouglasUP{ M.~R.~Douglas, I.~R.~Klebanov, D.~Kutasov,
J.~Maldacena, E.~Martinec and N.~Seiberg, ``A new hat for the c =
1 matrix model,'' arXiv:hep-th/0307195.
}

\lref\KutasovUA{ D.~Kutasov and N.~Seiberg, ``Noncritical
Superstrings,'' Phys.\ Lett.\ B {\bf 251}, 67 (1990).
}

\lref\MurthyES{ S.~Murthy, ``Notes on non-critical superstrings in
various dimensions,'' JHEP {\bf 0311}, 056 (2003)
[arXiv:hep-th/0305197].
}

\lref\McGreevyDN{ J.~McGreevy, S.~Murthy and H.~Verlinde,
``Two-dimensional superstrings and the supersymmetric matrix
model,'' JHEP {\bf 0404}, 015 (2004) [arXiv:hep-th/0308105].
}

\lref\GukovYP{ S.~Gukov, T.~Takayanagi and N.~Toumbas, ``Flux
backgrounds in 2D string theory,'' JHEP {\bf 0403}, 017 (2004)
[arXiv:hep-th/0312208].
}

\lref\KutasovQX{ D.~Kutasov, E.~J.~Martinec and N.~Seiberg,
``Ground rings and their modules in 2-D gravity with c <= 1
matter,'' Phys.\ Lett.\ B {\bf 276}, 437 (1992)
[arXiv:hep-th/9111048].
}

\lref\TakayanagiSM{ T.~Takayanagi and N.~Toumbas, ``A Matrix Model
Dual of Type 0B String Theory in Two Dimensions,''
arXiv:hep-th/0307083.
}

\lref\Seiberg{N.~Seiberg, unpublished.}


 \Title{ \rightline{hep-th/0502156} }
{\vbox{\centerline{Observations On The Moduli Space Of }
\centerline{Two Dimensional String Theory}}}
\medskip

\centerline{\it Nathan Seiberg}
\bigskip
\centerline{School of Natural Sciences} \centerline{Institute for
Advanced Study} \centerline{Einstein Drive, Princeton, NJ 08540}

\smallskip

\vglue .3cm

\bigskip
\noindent
 We explore the moduli space of the two dimensional fermionic
string with linear dilaton.  In addition to the known 0A and 0B
theories, there are two theories with chiral GSO projections,
which we call IIA and IIB. They are similar to the IIA and IIB
theories of ten dimensions, but are constructed with a different
GSO projection. Compactifying these theories on various twisted
circles leads to eight lines of theories.  Three of them, 0A on a
circle, super-affine 0A and super-affine 0B are known.  The other
five lines of theories are new.  At special points on two of them
we find the noncritical superstring.

\Date{02/05}

\newsec{Introduction}

There are several reasons to study low dimensional string
theories.  Since they are simpler than their higher dimensional
counterparts, they are often solvable.  The exact solution can
teach us about new phenomena which could be present also in more
generic situations.  In particular, these theories involve a
linear dilaton and are very similar to more complicated theories
with a linear dilaton like the NS5-brane theory.  It is important
to add that despite their simplicity, these theories are very rich
and exhibit many qualitative phenomena which are present in more
generic theories.

Here we will discuss the two dimensional theory with worldsheet
supersymmetry\foot{For early reviews of the bosonic version of
these theories see, e.g.\ \refs{\igor\ginsparg-\joe}.}. The target
space is parametrized by two coordinates $X$ and $\phi$. The
string metric is flat and the dilaton is linear in $\phi$.  We
will limit ourselves to theories which are translation invariant
in $X$. The extension to other theories, including various
orbifolds of our theories is straightforward. For most of our
discussion we will focus on the weak coupling end of the target
space $\phi \to -\infty$. Although various deformations of the
background like nonzero NS-NS ``tachyon'' fields or RR fluxes are
possible, they are negligible in that region and the worldsheet
theory can be analyzed there using free field methods.

There are four theories with noncompact $X$. Two of them, the 0A
and the 0B theories, are familiar (see, e.g.\
\refs{\DouglasUP,\TakayanagiSM} and references therein). The other
two theories, which are less known, can be called IIA and
IIB.\foot{Motivated by the noncritical superstring construction of
\KutasovUA\ (see also the review \KutasovPV), the spectrum of
these theories in noncompact space was guessed in \Seiberg\ and
mentioned in \refs{\MurthyES\McGreevyDN\GukovYP-\TakayanagiGE}.
Our discussion below will derive this spectrum, and will clarify
its connection to the noncritical superstring.}  The two type II
theories are similar to their ten dimensional analogs.  The
spectrum of the IIA theory is not chiral.  It consists of a single
Majorana fermion with its left and right moving components. The
IIB theory is chiral. Its spectrum has two Majorana Weyl fermions
of one chirality and a scalar of the opposite chirality, thus
cancelling the anomalies in a nontrivial way.  However, unlike
their ten dimensional counterparts, these theories do not have
spacetime supersymmetry.  More microscopically, it is important to
stress that the GSO projection in these two dimensional theories
is different than in ten dimensions. In section 2 we describe the
worldsheet construction of these theories and discuss the
relations between them.  We also present their ground rings and
the puzzles the new theories pose. Perhaps these puzzles, which
are reminiscent of black hole physics, could be resolved by
assuming that the semiclassical picture is not precise, and that
there are more asymptotic states in addition to the perturbative
quanta.

In section 3 we examine various twisted circle compactifications
of the four noncompact theories. This moduli space is similar to
its ten dimensional relative (see figure 1).  It includes eight
lines parametrized by the compactification radius $R$. Three of
these lines (lines $1 - 3$ in figure 1) are known from the study
of the type 0 theory. The other five lines (lines $4 - 8$ in
figure 1) are new. Two special points (marked with black squares
in figure 1), one on line 7 and the other on line 8 correspond to
the noncritical superstring of \KutasovUA.  Four other special
points (denoted by black circles in figure 1) have enhanced
nonabelian symmetry.

\medskip \ifig\figI{The moduli space of theories we consider.
The four corners of the square represent the four theories before
compactification.  They are labelled by 0B, 0A, IIB and IIA.  The
eight lines, labelled by $1 - 8$ represent different
compactifications.  They are labelled by their number in the text.
The points on each line represent compactifications with different
radii $R$.  Lines $1$, $4$, $7$ and $8$ interpolate between
different noncompact theories as $R$ varies between $ 0$ and $
\infty$.  Each point on lines $2$, $3$, $5$ and $6$ corresponds to
two different dual radii; the four points labelled by black
circles at the ends of these lines are the selfdual points. At
these points the theory has enhanced nonabelian symmetry.  The
points marked with black squares on lines $7$ and $8$ represent
the noncritical superstrings of \KutasovUA.}
    {\epsfxsize=0.6\hsize\epsfbox{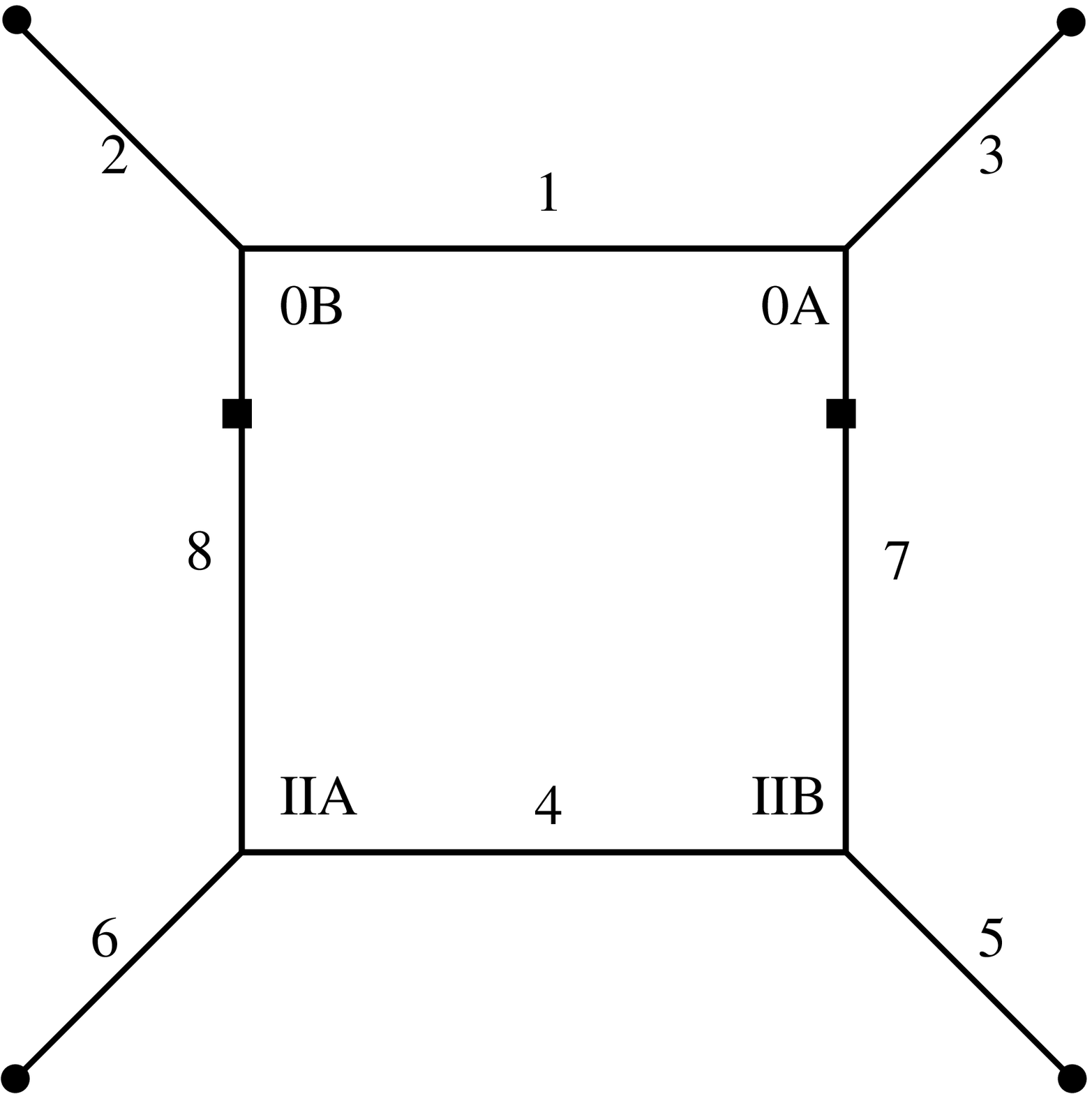}}

\subsec{Notations and Symmetries}

As preparation for our discussion, let us present our notations
and symmetries.

As a free worldsheet field, $X$ can be written as a sum of
worldsheet chiral components $X=x + \bar x$.  The fermionic
partners of $x$ and $\bar x$ are $\psi_x$ and $\bar \psi_x$.
Together with the superpartners of $\phi$ they can be bosonized to
worldsheet chiral scalars $H$ and $\bar H$. Throughout this paper
we set $\alpha'=2$.  If needed, dimensions can be restored by
multiplying every parameter with dimensions of length by
$\sqrt{\alpha'/2}$.

We use four discrete transformations which act on the various
fields as follows:
 \eqn\transformations{\eqalign{
 (-1)^{f_L} \qquad
 &\varphi \to \varphi + i \pi\qquad \qquad; \qquad \qquad H \to H
 +  \pi \cr
  (-1)^{F_L} \qquad
 &\varphi \to \varphi +  2\pi i\cr
  (-1)^{f_R} \qquad
 &\bar \varphi \to \bar \varphi + i \pi \qquad \qquad; \qquad \qquad
 \bar H \to \bar H +  \pi \cr
  (-1)^{F_R} \qquad
 &\bar \varphi \to \bar \varphi +  2\pi i}}
where $\varphi $ and $\bar\varphi$ are the left and right moving
bosonized superghosts.  $f_{L,R}$ are the left and right moving
worldsheet fermion numbers and $F_{L,R}$ are the left and right
moving spacetime fermion numbers. Since all our operators will be
constructed out of building blocks of the form $e^{-\varphi + i n
H}$ and $e^{-{\varphi \over 2} + i (n+{1\over 2} )H}$ with $n\in
\Bbb Z$ (and similarly for the right movers), all the operators in
\transformations\ square to one when they act on our vertex
operators.  Therefore, \transformations\ are $\Bbb Z_2$
transformations.

In addition to \transformations\ we will also be interested in the
two ``parity transformations''
 \eqn\parityt{\eqalign{
 P_{ws} \qquad & z \leftrightarrow \bar z\qquad ; \qquad \varphi
 \leftrightarrow \bar\varphi \qquad ; \qquad H \leftrightarrow
 \bar H \qquad ;
 \qquad x \leftrightarrow \bar x \cr
 P_{ts}\qquad &H \to - H\qquad ; \qquad \bar H \to - \bar H \qquad
 ; \qquad x \to - x \qquad ; \qquad \bar x \to - \bar x}}
$ P_{ws}$ is worldsheet parity and  $P_{ts}$ is the target space
parity.  Conjugation by them acts on \transformations.  Clearly,
 \eqn\pwscon{\eqalign{
 &(-1)^{F_{L,R}}=P_{ws} (-1)^{F_{R,L}}P_{ws} \cr
 &(-1)^{f_{L,R}}=P_{ws} (-1)^{f_{R,L}}P_{ws}
 }}
and using the fact that \transformations\ are $\Bbb Z_2$
transformations, we find \
 \eqn\ptscon{\eqalign{
 &(-1)^{F_{L,R}}=P_{ts} (-1)^{F_{L,R}}P_{ts}\cr
 &(-1)^{F_{L,R}+f_{L,R}}=P_{ts} (-1)^{f_{L,R}}P_{ts} }}

\newsec{Theories on ${\Bbb R}^2$}

There are four theories in noncompact space.  Each of them is
characterized by a $H=\Bbb Z_2\times \Bbb Z_2$ subgroup of the
four $\Bbb Z_2$ in \transformations, which acts trivially on all
the operators.  The remaining $G=\Bbb Z_2\times \Bbb Z_2$ acts as
symmetry. More precisely, the group $G$ is a quotient of the four
$\Bbb Z_2$ in \transformations\ by $H$; i.e.\ the elements of $G$
can be replaced by elements of $G$ times elements of $H$.

The four theories are related by orbifolding.  Orbifolding each
theory by various subgroups of its symmetry $G$ leads to other
theories. There are two well known subtleties in doing that.
First, in addition to the $G=\Bbb Z_2 \times \Bbb Z_2$ symmetry
the theories also have symmetries generated by the parity
transformations $P_{ws}$ and $P_{ts}$ which act on $G$ as in
\pwscon\ptscon.  Therefore, conjugation by $P_{ws}$ and $P_{ts}$
restricts the number of distinct orbifolds by subgroups of $G$.
Second, one has to be careful about the sign of the orbifold
projection in the twisted sectors. Normally, it is chosen by
consistency. Here, since we have an explicit realization in terms
of free fields, we simply have to impose the correct generators to
mod out by.  More explicitly, we can always combine the
orbifolding generator with an element of the group $H= \Bbb Z_2
\times \Bbb Z_2$ which acts trivially. Then we simply mod out by
the elements which act trivially in the resulting theory.

\subsec{0B}

Here the transformations $(-1)^{F_L+F_R} = (-1)^{f_L+f_R}=1$ act
trivially, and the symmetry $G$ is generated by $(-1)^{F_L} $ and
$(-1)^{f_L}$. The physical operators, ignoring for the moment
operators at special momenta are
 \eqn\zbspectrum{\eqalign{
 &T(p)=e^{-\varphi-\bar \varphi + i p (x + \bar x)+(1-|p|)\phi}
 \cr
 &C(p)= e^{-{\varphi \over 2}-{\bar\varphi \over 2} +{i
 \epsilon(p)\over 2} (H  + \bar H )+
 ip(x + \bar x)+(1-|p|)\phi}\qquad\qquad p \not =0\cr
 &C_\pm= e^{-{\varphi \over 2}-{\bar\varphi \over 2} \pm {i \over
 2} (H  +\bar H )+\phi}\cr
 &\epsilon(p) = {\rm Sign}(p)}}
Here and below, the absolute value of $p$ in the $\phi$ dependence
follows from the bound on the Liouville exponent \SeibergEB.
Linear combinations of $C_\pm$ are the zero modes of $C$ and its
dual. The propagating particles are the NS-NS ``tachyon'' $T$ and
the R-R scalar $C$. The symmetry $(-1)^{f_L}$ exchanges $C$ and
its dual, and therefore $C$ must be compact and its radius is the
selfdual radius \DouglasUP. The theory is also invariant under the
worldsheet and target space parities $P_{ws}$, $P_{ts}$.

In addition to the vertex operators \zbspectrum, the theory also
has special operators which exist only at certain discrete
momenta. For example,
 \eqn\curr{\eqalign{
 &j^0=e^{-\varphi } \psi_x \cr
 &\bar j^0=e^{-\bar\varphi } \bar \psi_x \cr}}
are worldsheet currents associated with a $U(1)\times U(1)$
symmetry. Their sum is the momentum $p$ in \zbspectrum\ and their
difference does not act on \zbspectrum\ (it is the winding
symmetry of the compactified theory).  Using these operators we
can construct the two dimensional version of the higher
dimensional dilaton/graviton
 \eqn\vop{V=j^0 \bar j^0 }

Other operators of interest are the ground ring operators
\refs{\WittenZD\BouwknegtYG\ImbimboIA\BouwknegtAM\BouwknegtVA
\Itoh\KutasovQX-\BershadskyUB}. These are dimension $(0,0)$ vertex
operators. The ring is freely generated by
 \eqn\zbringgen{
  R_\pm= e^{-{\varphi \over 2}-{\bar\varphi \over 2} \mp {i \over
  2} (H + \bar H ) \pm {i\over 2} (x + \bar x)-{\phi\over 2}}+
  \dots}
The vertex operators \zbspectrum\ form modules of the ring
\refs{\KutasovQX,\DouglasUP}.
 \eqn\zbmod{\eqalign{
 &R_+ C(p)=\cases{T(p+{1\over 2}) & $p>0$ \cr 0 & $p<0$} \cr
 &R_- C(p)=\cases{ 0 & $p>0$ \cr T(p-{1\over 2}) & $p<0$ } \cr
 &R_+ T(p)=\cases{-p^2 C(p+{1\over 2}) & $p>0$ \cr 0 & $p<0$} \cr
 &R_- T(p)=\cases{ 0 & $p>0$ \cr -p^2 C(p-{1\over 2}) & $p<0$ }}}
(the factors of $-p^2$ in the action on $T(p)$ arise from picture
changing.)  In all these modules we have the relation $R_+R_-=0$.
When a zero momentum tachyon is turned on by adding $\mu T(0)$ to
the worldsheet Lagrangian, there are still no relations in the
ring, but the module relation is deformed to
\refs{\KutasovQX,\DouglasUP}
 \eqn\modrelzb{R_+R_-= \mu}
In the matrix model description of these theories, the fields $T$
and $C$ are interpreted as ripples on the Fermi sea, and the
relation \modrelzb\ is interpreted as the shape of the Fermi sea.

Besides turning on the tachyon operator $\mu T(0)$, we can also
preserve translation invariance while turning on the RR operators
 \eqn\rrdefzb{q_+ C_+ + q_-C_- = q(C_++C_-) + \tilde q(C_+-C_-)}
It is not easy to use worldsheet methods to compute the
consequences of such deformations, but this should be doable using
matrix models.

Finally, let us mention the action of $G=\Bbb Z_2\times \Bbb Z_2$
on these deformations
 \eqn\symdefzb{\eqalign{
 &(-)^{F_L} \qquad \qquad q_\pm \to - q_\pm \cr
 &(-1)^{f_L} \qquad \qquad \mu \to -\mu\qquad; \qquad q_\pm \to
 \pm q_\pm}}

\subsec{0A}

An orbifold of the 0B theory by its symmetry $(-1)^{F_L}$ leads to
the 0A theory. More precisely, in the 0A theory the
transformations $(-1)^{F_L+F_R} = (-1)^{f_L+f_R+F_L}=1$ act
trivially, and $(-1)^{F_L} $ and $ (-1)^{f_L}$ generate the
symmetry $G$. The spectrum includes
 \eqn\zaspectrum{\eqalign{
 &T(p)=e^{-\varphi-\bar \varphi + i p (x + \bar x) + (1-|p|) \phi}
 \qquad\qquad p\in \Bbb R\cr
 &F^\pm= e^{-{\varphi \over 2}-{\bar\varphi \over 2} \pm {i\over
 2}  (H  - \bar H )-\phi}}}
Here $F^\pm$ are the two RR fluxes of the theory.  They are the
zero momentum modes of two different gauge fields.  The only
propagating particle is the ``tachyon'' $T$.  As in the 0B theory,
the theory is also invariant under the two parity transformations
$P_{ws}$ and $P_{ts}$ \parityt.

As in the 0B theory we have the currents \curr\ and the
dilaton/graviton \vop. The ground ring, however, is a quotient of
the 0B ground ring by $(-1)^{F_L}$; i.e.\ it includes only the
even powers of $R_\pm$. It is generated by $O_+=R_+^2$,
$O_-=R_-^2$ and $O_0=R_+R_-$ with the relation $O_+O_-=O_0^2$.
Clearly, the tachyons $T(p)$ are in modules of the ring with the
relation $O_0=0$.

Again, as in the 0B theory there are three translation invariant
deformations
 \eqn\defza{\mu T(0)+ q_+ F^+ + q_-F^- = \mu T(0)+ q(F^++F^-)
 + \tilde q(F^+-F^-)}
and the action of $G=\Bbb Z_2\times \Bbb Z_2$ on the deformations
is as in \symdefzb.  The deformations \defza\ deform the module
relation $O_0=0$ to $O_0=\mu$.

\subsec{IIB}

An orbifold of the 0B theory by its symmetry $(-1)^{f_L}$ leads to
the IIB theory.  More precisely, here the transformations
$(-1)^{f_L+f_R} = (-1)^{f_L+F_L+F_R}=1$ act trivially, and $G$ is
generated by $(-1)^{F_L} $ and $ (-1)^{F_R}$.  The spectrum
includes
 \eqn\tbspectrum{\eqalign{
 & \Psi_-(p)=e^{-{\varphi \over 2}-\bar \varphi-i{H \over 2}+
 ip(x+ \bar x)+ (1-|p|) \phi} \qquad\qquad p\le 0\cr
 &\bar\Psi_-(p)=e^{-\varphi-{\bar\varphi \over 2} -i {\bar H
 \over 2}+ ip(x+ \bar x)+ (1-|p|) \phi}\qquad\qquad p \le 0\cr
 &C_+(p)= e^{-{\varphi \over 2}-{\bar\varphi \over 2} +
 {i\over 2} (H  + \bar H )+ ip(x + \bar x)+ (1-|p|) \phi}
 \qquad\qquad p\ge 0}}
The spectrum of particles is a left moving boson $C_+$ and two
right moving Majorana Weyl fermions $\Psi_-(p)$ and
$\bar\Psi_-(p)$.  The chiral nature of the spectrum originates
from the absolute value of $p$ in the exponent of $\phi$, which
arises from the bound of \SeibergEB.  The theory has only one
RR-flux, $C_+(0)$.  The theory is invariant under the worldsheet
parity transformation $P_{ws}$ but not under $P_{ts}$. Clearly,
$(-1)^{F_{L,R}}$ are related by conjugation by this symmetry.
Here, an isomorphic theory (IIB') can be obtained by the action of
the target space parity operation $P_{ts}$.

It is important that the projections in the R-NS and NS-R sectors
are opposite to the ten dimensional ones, where the projections
are $(-1)^{f_L}=(-1)^{f_R}=1$.  Had we used such a projection,
there would not have been any fermions and the spectrum would have
included only the left moving boson $C_+$. This spectrum is
anomalous and the theory is likely to be inconsistent.

Unlike the type 0 theory, here there is only one possible
translation invariant deformation -- the RR flux $q_+ C_+(0)$.

As in the 0B theory, we have the operators \curr\vop. The ground
ring, however, is a quotient of the 0B ground ring by
$(-1)^{f_L}$.  It is freely generated by $R_-$ and $O_+=R_+^2$.
The vertex operators \tbspectrum\ are again in modules of the ring
 \eqn\tbmod{\eqalign{
 &R_- \Psi_-(p)=i p \bar \Psi_-(p-{1\over 2})\cr
 &R_-  \bar \Psi_-(p)=i p \Psi_-(p-{1\over 2})\cr
 &R_- C_+(p)=0 \cr
 &O_+ \Psi_-(p)=0\cr
 &O_+ \bar \Psi_-(p)=0\cr
  &O_+ C_+(p)=-(p+{1\over 2})^2 C_+(p+1)}}
(The factor of $i p$ in the RHS arises from picture changing.)  In
all these modules we have $R_-O_+=0$.  By analogy to the known 0A
and 0B matrix models, we guess that this theory also has a matrix
model, where equation \tbmod\ means that $\Psi_-$ and $ \bar
\Psi_-$ correspond to ripples on one half of a Fermi surface,
while $C_+$ describes ripples on the other half.

What is the S-matrix of this theory?  The incoming particles are
the fermions $\Psi_-$ and $\bar \Psi_-$, and the outgoing
particles are the chiral bosons $C_+$.  Denote the number of
$\Psi_-$ and $\bar \Psi_-$ incoming quanta by $n_-$ and $\bar n_-$
and the number of outgoing $C_+$ quanta by $n_+$. Then, the
discrete symmetries generated by $(-1)^{F_{L,R}}$ restrict the
S-matrix, to obey $n_- + \bar n_- \in 2 \Bbb Z$ and $n_- + n_+ \in
2 \Bbb Z$.  If we turn on background $C_+(0)$, the second
restriction is clearly lifted, but the first restriction remains.

This is very peculiar.  Consider a process with an odd number of
incoming quanta carrying an odd fermion number.  Since all the
outgoing quanta are bosons, $(-1)^{F_L+F_R}$ cannot be conserved.
In other words, this S-matrix is not unitary! This means that the
theory should have more asymptotic states in addition to the
quanta we mentioned here. For example, it is possible that a
coherent state of $C_+$ quanta carries the necessary fermion
number.  An explicit example of this phenomenon which is very
similar to our case is given in \refs{\CallanUB,\PolchinskiMY}.
Such behavior of the S-matrix, where nonunitarity in perturbative
calculations is fixed in the full theory, is reminiscent of black
hole physics, and is therefore worthy of further study.  A
possible interesting direction is to look for a matrix model of
this system (for first attempts see
\refs{\McGreevyDN\GukovYP-\TakayanagiGE}).

\subsec{IIA}

An orbifold of the IIB theory by $(-1)^{F_L}$ (or alternatively,
an orbifold of the 0A theory by $(-1)^{f_L}$) leads to the IIA
theory.  More precisely, here the transformations $(-1)^{F_L+f_R}
= (-1)^{f_L+f_R+F_R}=1$ act trivially, and $(-1)^{F_L} $ and
$(-1)^{F_R}$ generate the symmetry $G$.  The spectrum includes
 \eqn\taspectrum{\eqalign{
 &\Psi_-(p)=e^{-{\varphi \over 2}-\bar\varphi -i{H \over 2}+
 ip(x+ \bar x)+ (1-|p|) \phi} \qquad\qquad p\le 0\cr
 &\Psi_+(p)=e^{-\varphi-{\bar\varphi \over 2} +i {\bar H \over 2}+
 ip(x+ \bar x)+ (1-|p|) \phi}\qquad\qquad p\ge 0\cr
 &F^+= e^{-{\varphi \over 2}-{\bar\varphi \over 2} + {i\over 2} (H
 - \bar H )+  \phi}}}
Here $F^+$ is the single RR flux of the theory.  It is the field
strength of a gauge field.  The spectrum of particles is a single
left moving and right moving Majorana fermion $ \Psi_\pm$.
Consider the parity transformations \parityt.  This theory is
invariant only under the $\Bbb Z_2$ generated by the combined
operation $P_{ws}P_{ts}$. Using \pwscon\ptscon, the two symmetries
$(-1)^{F_{L,R}}$ which we mentioned above are related by
conjugation by $P_{ws}P_{ts}$. Acting with either $P_{ws}$ or
$P_{ts}$ leads to another isomorphic theory (IIA').

As in the IIB theory, the projection in the R-NS and the NS-R
sectors is opposite to the ones in ten dimensions, where
$(-1)^{f_L} =(-1)^{f_R+F_R}=1$.  That projection would not have
allowed any operators from the R-NS and NS-R sectors in two
dimensions and presumably would have resulted in an inconsistent
theory.

Here, as in the IIB theory, there is only one translation
invariant deformation of the theory -- the RR flux $q_+ F^+$.

Again, we have the operators \curr\vop. The ground ring is a
quotient of the IIB ground ring by $(-1)^{F_L}$ (or equivalently
of the 0A ground ring by $(-1)^{f_L}$).  It is freely generated by
$O_+=R_+^2$ and $O_-=R_-^2$.  The fermions $\Psi_-(p)$ are in
modules of this ring with the module relation $O_+=0$, while the
fermions $\Psi_+(p)$ are in modules of this ring with the module
relation $O_-=0$.  As in the IIB discussion, we expect a matrix
model of this system to exist, where $\Psi_\pm(p)$ correspond to
ripples on different halves of the Fermi surface.

As in the IIB theory, the S-matrix of this theory is interesting.
The incoming quanta are the fermions $\Psi_-$ and the outgoing
quanta are the fermions $\Psi_+$.  In the absence of background RR
flux $q_+ F^+$ the S-matrix of these quanta is not unitary without
possible coherent states.  Unlike the IIB theory, there is a
possible unitary scattering when RR flux is turned on. Clearly,
this S-matrix should be further explored, perhaps by finding a
matrix model for this system.

\newsec{The theories on $\Bbb R \times S^1$}

These four theories can be compactified on a circle with various
twists by their symmetry elements.  We will limit ourselves to
compactifications without twists by the two parity transformations
\parityt, and will also not consider possible asymmetric
orbifolds.  A priori each theory can be twisted by any of the four
elements of its $G=\Bbb Z_2 \times \Bbb Z_2$ symmetry. However,
since in all our cases two of these elements are related by
conjugation by a certain parity symmetry of the theory, only three
of these compactifications are distinct. Finally, T-dualities
relate these 12 compactified theories leaving only eight
independent ones.

All the theories have the currents \curr\ and the dilaton/graviton
operator \vop, which changes the radius $R$.  We will not list
them explicitly except in special cases.  There are also various
discrete states and ring elements, most of them will not be
discussed.

The torus partition function of these theories are divergent both
in the weak coupling and the strong coupling regions.  But its
value per unit length is finite.  This value is not easy to
compute because of subtleties in the odd spin structures. In
\DouglasUP, where our first three examples were studied in detail,
it was argued that the torus partition function can be obtained by
summing over the various physical modes using $\zeta$-function
regularization; i.e. $\sum n \to -{1\over 12}$, $\sum 2n \to
-{2\over 12}$ and $\sum 2n+1 \to +{1\over 12}$.  In doing it we
should remember to include target space fermions with an extra
minus sign. We will use this procedure below.

The various compactifications are (in order not to clutter the
equations, we suppressed the dependence on $\phi$):

\subsec{Line 1: 0A/0B on a circle}

0A compactified on a circle of radius $R$ (without a twist) is
T-dual to 0B on ${2\over R}$.  The spectrum is
 \eqn\onespectrum{\eqalign{
 &T(p)=e^{-\varphi-\bar \varphi + i {p\over R} (x + \bar x)}\qquad
 \qquad p=0,\pm 1,\pm 2 ... \cr
 &\tilde T(w)=e^{-\varphi-\bar \varphi + i {w R \over 2} (x- \bar
 x)}\qquad \qquad w=\pm 1,\pm 2, ... \cr
 &\tilde C(w)= e^{-{\varphi \over 2}-{\bar\varphi \over 2} +
 {i \epsilon(w) \over 2} (H  - \bar H )+ i{w R \over 2}(x - \bar x)}
 \qquad \qquad w =\pm 1,\pm 2 ...\cr
 &F^\pm =\tilde C_\pm= e^{-{\varphi \over 2}-{\bar\varphi \over 2}
 \pm {i \over 2} (H  - \bar H )}}}
Here we denoted by tilde the fields of the noncompact T-dual
theory.

The torus partition function is
 \eqn\onetor{{\cal Z}_1= -{1\over 12} \left({1\over R} +
 R\right)}

In addition to the three deformations of the 0A or 0B theories on
${\Bbb R}$, here we can preserve one of the shift symmetries
around the circle by turning on operators with either nonzero
momentum or nonzero winding.  If such an operator is $T$ or
$\tilde T$, it leads to a Sine-Liouville interaction on the
worldsheet.  We will not explore it here.

\subsec{Line 2: Super-affine 0B}

This is a compactification of the 0B theory such that as we move
around a circle of radius $R$ we twist by $(-1)^{F_L}$; i.e.\ we
mod out the theory on a circle with radius\foot{It is common in
the literature to denote by $R$ the radius of this circle; i.e.\
$R_{there}=2R_{here}$.} $2R$ by $(-1)^{F_L} e^{i\pi p}$. The
spectrum is
 \eqn\twospectrum{\eqalign{
 &T(p)=e^{-\varphi-\bar \varphi + i {p\over R} (x + \bar
 x)}\qquad \qquad p=0, \pm 1, \pm 2,...\cr
 &\tilde T(w)=e^{-\varphi-\bar \varphi + i {w R \over 2}(x- \bar x)}\qquad
 \qquad w=\pm 2, \pm 4...\cr
 &C(p)= e^{-{\varphi \over 2}-{\bar\varphi \over 2} +
 {i \epsilon(p) \over 2} (H  + \bar H )+ i{p\over R}
 (x + \bar x)}\qquad \qquad p =\pm
{1\over 2}, \pm {3\over 2} ... \cr
 &\tilde C(w)= e^{-{\varphi \over 2}-{\bar\varphi \over 2}
 +{i \epsilon(w) \over 2} (H  - \bar H )+
 i{w R \over 2}(x - \bar x)}\qquad \qquad w=\pm 1, \pm 3,...\cr
 }}

These theories are selfdual under the transformation $R\to {1\over
R}$.  At the selfdual radius $R=1$ the operators \curr\ are
accompanied by two more left moving and two more right moving
worldsheet currents leading to $SU(2) \times SU(2)$ symmetry:
 \eqn\sutc{\eqalign{
 &j^\pm = e^{-\varphi \pm i x} \cr
 & j^0 = e^{-\varphi}\psi_x \cr
 &\bar j^\pm = e^{-\bar \varphi \pm i \bar x} \cr
 &\bar j^0 = e^{-\bar \varphi}\bar \psi_x}}
All the physical operators are in representations of this
symmetry.  For example, the dilaton/graviton operator \vop, the
operators $T(p=\pm1)$, $\tilde T(w=\pm2)$ in \twospectrum\ and
four other operators form the $({\bf 1,1})$ representation.  They
can be written as $j^a \bar j^b$ with $a,b=\pm,0$. Similarly,
$(C(p=\pm {1\over 2}), \tilde C(w=\pm 1))$ are in the $({\bf {1
\over 2},{1 \over 2}})$ representation.  Away from the selfdual
point the operators $j^\pm$ and $\bar j^\pm$ do not deform to
physical operators and they simply disappear.

The torus partition function is
 \eqn\twotor{{\cal Z}_2= -{1\over 24} \left({1\over R} +
 R\right)}

Finally, note that the twist around the circle eliminates the zero
momentum deformations \rrdefzb.  This means that this
compactification is possible only when $q_+=q_-=0$.

 \subsec{Line 3: Super-affine 0A}

This is a compactification of the 0A theory such that as we move
around a circle of radius $R$ we twist by $(-1)^{F_L}$; i.e.\ we
mod out the theory on a circle with radius $2R$ by $(-1)^{F_L}
e^{i\pi p}$. The spectrum is
 \eqn\threespectrum{\eqalign{
 &T(p)=e^{-\varphi-\bar \varphi + i {p\over R} (x + \bar
 x)}\qquad \qquad p=0,\pm 1,\pm 2 ...\cr
 &\tilde T(w)=e^{-\varphi-\bar \varphi + i {w R \over 2}(x- \bar x)}\qquad
 \qquad w=\pm 2,\pm 4 ...
 }}
These theories are selfdual under the transformation $R\to {1\over
R}$, and at the selfdual radius $R=1$ have the $SU(2)\times SU(2)$
currents $j^{\pm, 0}$, $\bar j^{\pm,0}$ of \sutc\ and the vertex
operators $j^a \bar j^b$. The torus partition function is
 \eqn\threetor{{\cal Z}_3= -{1\over 12} \left({1\over R} +
 R\right)}
and as in the 0B super-affine line, this compactification is
possible only when $q_+=q_-=0$.

\subsec{Line 4: IIA/IIB on a circle}

IIA compactified on a circle with radius $R$ (without a twist) is
T-dual to IIB on ${2\over R}$.  The physical operators are
 \eqn\fourspectrum{\eqalign{
 & \Psi_-(p)=e^{-{\varphi \over 2}-\bar \varphi-i{H \over 2}+
 i{p\over R}(x+ \bar x)} \qquad\qquad p =0,-1,-2 ...\cr
 &\tilde \Psi_-(w)=e^{-{\varphi \over 2}-\bar \varphi-i{H \over
 2}+ i{w R \over 2}(x- \bar x)} \qquad\qquad w=-1,-2 ...\cr
 &\Psi_+(p)=e^{-\varphi-{\bar\varphi \over 2} + i {\bar H \over
 2}+ i{p\over R}(x+ \bar x)}\qquad\qquad p =0,1,2 ...\cr
  &\tilde {\bar\Psi}_-(w)=e^{-\varphi-{\bar\varphi \over 2} +
  i {\bar H \over 2}+ i{w R \over 2}(x- \bar x)}\qquad\qquad w =-1,-2 ...\cr
 &\tilde C_+(w)= e^{-{\varphi \over 2}-{\bar\varphi \over 2} +
 {i \over 2}(H - \bar H )+ i{w R \over 2}(x - \bar x)}\qquad\qquad
 w=0,1,2 ...}}
The operator $\tilde C(0)=F$ is the RR flux.  It can be added to
the worldsheet Lagrangian without breaking the two translation
symmetries of the compactification.  The torus partition function
is
 \eqn\fourtor{{\cal Z}_4= +{1\over 24} \left({2\over R} +
 {R \over 2} \right)}

\subsec{Line 5: Super-affine IIB}

Here the IIB theory is compactified on a circle of radius $R$
twisted by $(-1)^{F_L}$.
It is selfdual with $R\to {1\over R}$.  The spectrum is
 \eqn\fivespectrum{\eqalign{
 & \Psi_-(p)=e^{-{\varphi \over 2}-\bar \varphi-i{H \over 2}+
 i{p\over R}(x+ \bar x)} \qquad\qquad p =-{1\over 2},-{3\over 2},
 ...\cr
 &\tilde \Psi_+(w)=e^{-{\varphi \over 2}-\bar \varphi+i{H \over
 2}+ i{w R \over 2}(x- \bar x)} \qquad\qquad w=1,3, ...\cr
 &\bar \Psi_-(p)=e^{-\varphi-{\bar\varphi \over 2} - i {\bar H
 \over 2}+ i{p\over R}(x+ \bar x)}\qquad\qquad p =0,-1,-2 ...\cr
 &\tilde {\bar\Psi}_+(w)=e^{-\varphi-{\bar\varphi \over 2}
  -i {\bar H \over 2}+ i{w R \over 2}(x- \bar x)}\qquad\qquad w
  =2,4,  ...\cr
 & C_+(p)= e^{-{\varphi \over 2}-{\bar\varphi \over 2}
 +{i\over 2}(H  + \bar H )+ i{p \over R}(x +\bar x)}
 \qquad\qquad p={1\over 2},{3\over 2},... \cr
 &\tilde C_-(w)= e^{-{\varphi \over 2}-{\bar\varphi \over 2}
 -{i\over 2}(H  - \bar H )+i{w R \over 2}(x - \bar x)}
 \qquad\qquad w=-1,-3,...
 }}
As in the super-affine lines of the type 0 theories, at the
selfdual radius $R=1$ we have enhanced symmetry.  But this time
the symmetry is $SU(2)\times U(1)$ with the currents $j^{\pm, 0}$
and $\bar j^0$ (without $\bar j^{\pm}$) of \sutc.  The spectrum is
in representations of this symmetry.  For example, $( \Psi_-(p=
-{1\over 2}) , \tilde \Psi_+(w= 1))$ and $( C_+(p= {1\over 2}) ,
\tilde C_-(w= -1))$ are doublet with $U(1)$ charges $-{1\over 2}$
and $+{1\over 2}$ respectively.. Using the currents we also find
the three NS-NS vertex operators $j^a \bar j^0$.

The torus partition function is
 \eqn\fivetor{{\cal Z}_5= +{1\over 24} \left({1\over R} +
 R\right)}

\subsec{Line 6: Super-affine IIA}

Here the IIA theory is compactified on a circle of radius $R$
twisted by $(-1)^{F_L}$. It is selfdual with $R\to {1\over R}$.
The spectrum is
 \eqn\sixspectrum{\eqalign{
 & \Psi_-(p)=e^{-{\varphi \over 2}-\bar \varphi-i{H \over 2}+
 i{p\over R}(x+ \bar x)} \qquad\qquad p =-{1\over 2},-{3\over 2},
 ...\cr
 &\tilde \Psi_+(w)=e^{-{\varphi \over 2}-\bar \varphi+i{H \over
 2}+ i{w R \over 2}(x- \bar x)} \qquad\qquad w=1,3, ...\cr
 &\Psi_+(p)=e^{-\varphi-{\bar\varphi \over 2} + i {\bar H \over
 2}+ i{p\over R}(x+ \bar x)}\qquad\qquad p =0,1,2 ...\cr
  &\tilde {\bar\Psi}_-(w)=e^{-\varphi-{\bar\varphi \over 2}
  +i {\bar H \over 2}+ i{w R \over 2}(x- \bar x)}\qquad\qquad w =-2,-4,
  ...
 }}
As in the super-affine IIB line, at the selfdual radius $R=1$ we
have the symmetry $SU(2)\times U(1)$ with the currents $j^{\pm,
0}$ and $\bar j^0$ (without $\bar j^{\pm}$) of \sutc\ and the
three vertex operators $j^a \bar j^0$.

The torus partition function is
 \eqn\sixtor{{\cal Z}_6= +{1\over 48} \left({1\over R} +
 R\right)}

\subsec{Line 7: IIB on a thermal circle}

Here we compactify the IIB theory on a thermal circle of radius
$R$; i.e.\ we twist with $(-1)^{F_L+F_R}$ (or equivalently, with
$(-1)^{f_L}$). It is T-dual to 0A twisted by $(-1)^{f_L}$ on
$2\over R $.  The physical operators are
 \eqn\sevenspectrum{\eqalign{
 &\tilde T(w)= e^{-\varphi-\bar\varphi+i{w R \over 2}(x-\bar x)}
 \qquad\qquad w=\pm 1, \pm 3 , ... \cr
 &\Psi_-(p)=e^{-{\varphi\over 2}-\bar\varphi - i {H \over 2}
 +i{p\over R}(x+\bar x)}
 \qquad\qquad p=-{1\over 2}, -{3\over 2}, ...\cr
 &\bar\Psi_-(p)=e^{-\varphi-{\bar\varphi\over 2} - i {\bar H \over
 2} +i{p\over R}(x+\bar x)}
 \qquad\qquad p=-{1\over 2}, -{3\over 2}, ...\cr
 &C_+(p)= e^{-{\varphi\over 2}-{\bar\varphi\over 2} + {i\over 2}(H
 +\bar H) +i{p\over R}(x+\bar x)}
 \qquad\qquad p=0,1,2,...\cr}}
For $R=1$ we have a phenomenon similar to \sutc\ but in the Ramond
sector.  Here we find the worldsheet currents ${\cal
S}_+=e^{-{\varphi \over 2} + i {H\over 2} + i x}$ and $\bar {\cal
S}_+=e^{-{\bar\varphi \over 2} + i {\bar H\over 2} + i \bar x}$,
which lead to fermionic target space symmetries with the algebra
$\{{\cal S_+, S_+}\}=\{{\cal \bar S_+, \bar S_+}\}=0$. Using these
operators we find the ``gravitino'' vertex operators ${\cal S}_+
e^{-\bar \varphi} \bar \psi_x$ and ${\cal \bar S}_+ e^{- \varphi}
\psi_x$.  Note that $C_+(p=1)= {\cal S_+ \bar S_+}$. This theory
coincides with the construction of \KutasovUA. As with \sutc, away
from $R=1$ the operators ${\cal S}_+$ and ${\cal \bar S}_+$ do not
deform to physical operators.

The torus partition function is
 \eqn\seventor{{\cal Z}_7= -{1\over 12} \left({1\over R}
 -{R\over 2}\right)}

\subsec{Line 8: IIA on a thermal circle}

Here we compactify the IIA theory on a thermal circle of radius
$R$; i.e.\ we twist with $(-1)^{F_L+F_R}$ (or equivalently, with
$(-1)^{f_L}$). It is T-dual to 0B twisted by $(-1)^{f_l}$ on
$2\over R $.  The physical operators are
 \eqn\eightspectrum{\eqalign{
 &\tilde T(w)=e^{-\varphi-\bar\varphi+i{w R \over 2}(x-\bar x)}
 \qquad\qquad w=\pm 1, \pm 3 , ...\cr
 &\Psi_-(p)=e^{-{\varphi\over 2}-\bar\varphi - i {H\over 2}
 +i{p\over R}(x+\bar x)}
 \qquad\qquad p=-{1\over 2}, -{3\over 2}, ...\cr
 &\Psi_+(p)=e^{-\varphi-{\bar\varphi\over 2} + i {\bar H\over 2}
 +i{p\over R}(x+\bar x)}
 \qquad\qquad p={1\over 2}, {3\over 2}, ...\cr
 &\tilde C_-(w)=e^{-{\varphi\over 2} -{\bar\varphi\over 2}
 - {i\over 2}(H-\bar H) +i {w R \over 2}(x-\bar x)}
 \qquad\qquad w=-1, -3, ...\cr
 &\tilde C_+(w)=e^{-{\varphi\over 2}-{\bar\varphi\over 2} +
 {i\over 2}(H-\bar H) +i {w R \over 2}(x-\bar x)}
 \qquad\qquad w=0, 2, ...\cr} }
For $R=1$ there are worldsheet currents ${\cal S}_+=e^{-{\varphi
\over 2} + i {H\over 2} + i x}$ and $\bar {\cal
S}_-=e^{-{\bar\varphi \over 2} - i {\bar H\over 2} - i \bar x}$
which lead to fermionic target space symmetries with the algebra
$\{{\cal S_+, S_+}\}=\{{\cal \bar S_-, \bar S_-}\}=0$. Using these
operators we find the ``gravitino'' vertex operators ${\cal S}_+
e^{-\bar \varphi} \bar \psi_x$ and ${\cal \bar S}_- e^{- \varphi}
\psi_x$.  Note that $C_+(w=2)= {\cal S _+\bar S_-}$. This theory
coincides with the construction of \KutasovUA. As with \sutc, away
from $R=1$ the operators ${\cal S}_+$ and ${\cal \bar S}_-$
disappear.

The torus partition function is
 \eqn\eighttor{{\cal Z}_8= -{1\over 24} \left({1\over R} -
 {R\over 2}\right)}

\newsec{The branes}

Here we consider the analog of the FZZT branes
\refs{\fzz,\Teschner} of our four theories on ${\Bbb R}^2$. We
will use the fermionic string version of the FZZT branes, which
were discussed in
\refs{\FukudaBV\korean\KlebanovWG\SeibergNM-\SeibergEI}.

Since we have limited ourselves to the linear dilaton theory
without a ``Liouville wall'', the results are going to be
meaningful only when the branes dissolve at the weak coupling
region.  For the type 0 theories we can find the branes by
considering the branes of the type 0 theory for nonzero $\mu$ in
the limit $\mu \to 0$ with fixed $\mu_B$.  These branes are
expressed in terms of the parameter $\sigma$.  From the worldsheet
point of view this is the Dirichlet boundary condition on the
Backlund field \SeibergNM, and from the matrix model point of
view, where $\mu_B$ is the eigenvalue coordinate, $\sigma$ is
essentially the ``time of flight variable.''  For the different
branes $\mu_B \sim \sqrt{|\mu|} \cosh\sigma$ or $\mu_B \sim
\sqrt{|\mu|} \sinh\sigma$.  Therefore, we are interested in the
limit $\mu\to 0$, $|\sigma| \to \infty $ with $\mu_B \sim
\sqrt{|\mu|} e^{|\sigma|}$ fixed.  In this limit the expressions
for the boundary states simplify.

Consider first the 0B theory.  Here we find 6 branes.  Two of them
are RR neutral and satisfy Neumann boundary conditions in
Euclidean time ($X= x + \bar x$).  They differ by the boundary
conditions of the supercharge $Q=\eta \bar Q$ with $\eta=\pm 1$.
There are also four RR charged branes labelled by $\eta=\pm 1$ and
$\xi=\pm1$ ($\xi$ distinguishes branes and anti-branes), which
have Dirichlet boundary conditions in Euclidean time. Taking $\mu
\to 0$, the boundary states are
 \eqn\zbb{\eqalign{
 |\mu_B, 0, \eta\rangle&= \sqrt{2} \int_0^\infty  dp\ \mu_B^{-2ip}
 A_{NS}(p) |NS,p, \partial X=0 , \eta\rangle \cr
 |\mu_B, X_0, \xi , \eta\rangle&=\int_0^\infty  dp\ \mu_B^{-2ip}
 \left( A_{NS}(p) | NS, p, X=X_0 , \eta\rangle  + \xi  A_{R}(p)
 | R , p, X=X_0, \eta \rangle \right)
 }}
where $p$ is the Liouville momentum, $A_{NS}(p)$ and $A_{R}(p)$
are momentum dependent functions, whose form will not be important
in this discussion, $|NS, p, \partial X=0 , \eta\rangle$ and $|
NS, p, X=X_0 , \eta\rangle$ are NS Ishibashi state with Liouville
momentum $p$ and Neumann or Dirichlet boundary conditions for $ X
$. They include the boundary states for the ghosts.  Similarly, $|
R , p, X=X_0, \eta \rangle $ are R Ishibashi states.  The relative
factor of $\sqrt 2$ between the neutral and charged branes is
common.  It is familiar from the study of Cardy states in the
Ising model.  It guarantees that the lowest open string tachyon
appears once on the brane. Note that the only dependence on
$\mu_B$ is through the simple factor $\mu_B^{-2ip}$ which
reproduces KPZ scaling (recall that $\mu=0$ and hence the only
dimensional parameter is $\mu_B$).

Clearly, the symmetry $(-1)^{F_L} $ leaves the neutral brane
invariant and exchanges the charged brane with its anti-brane,
$\xi \to -\xi$.  The symmetry $(-1)^{f_L}$ changes the sign of the
boundary conditions of the supercharge and maps $\eta\to -\eta$.

The actual branes are obtained by exponentiating the boundary
states \zbb.  Since in the asymptotic region the theory has an
$SU(2) \times SU(2)$ symmetry associated with the RR scalar
$C=C_++C_-$, the branes are in representations of this symmetry.
As always, the charge of the brane is localized near the point it
dissolves $(X=X_0, \phi \sim \log |\mu_B|)$.  Clearly, the neutral
branes are in singlets, and the four charged branes are in $({\bf
{1\over 2}, {1\over 2}})$.\foot{The fact that they carry charges
$\pm {1\over 2}$ is consistent with the discussion in \SeibergEI\
about the charges of such branes.} These latter branes can be
thought of as insertions of target space vertex operators
 \eqn\zbv{e^{{i \xi \over \sqrt{2} }( \eta C_+ + C_-)}}
times a function of $T$ at the point the brane dissolves.

In the 0A theory the neutral and charged branes are reversed.
There are two neutral branes which are localized in $X$ and four
charged branes which have Neumann boundary conditions in $X$.  The
latter are charged under the two gauge fields of the type 0A
theory, and their charges are localized near the point where the
branes dissolve.

The two type II theories are obtained by moding out the type 0
theories by $(-1)^{f_L}$.  Therefore, the branes in the type IIB
theory are
 \eqn\tbb{\eqalign{
 |\mu_B, 0 \rangle&= {1\over \sqrt {2}} \left( |\mu_B, 0, \eta=+1
 \rangle - |\mu_B,0, \eta=-1\rangle \right) \cr
  |\mu_B, X_0, \xi \rangle&= {1\over \sqrt {2}} \left( |\mu_B,X_0,
 \xi, \eta=+1 \rangle - |\mu_B,X_0, \xi, \eta=-1\rangle \right)
 \cr
 }}
Here the relative sign of the two branes of the type 0B theory is
such that the closed string NS-NS tachyon is projected out.
Therefore, the only physical NS-NS states which couple to these
branes are discrete states like the dilaton/graviton operator
$V=e^{-\varphi - \bar \varphi} \psi_x \bar \psi_x$.  Similarly,
only one chirality of the RR field, $C_+$ couples to the brane.
Finally, the normalization factor ${1\over \sqrt {2}}$ is such
that there is only one open string tachyon on the brane.  Because
of this normalization, the analog of \zbv\ is
 \eqn\tbv{e^{i \xi   C_+ }}
i.e.\ it is like a free fermion.  This suggests that the
projection from 0B to IIB does not simply remove $C_-$, but it
also changes the radius of $C_+$.

The branes of the type IIA theory are similar.  There is a single
neutral brane which is localized in $X$ and a charged brane and
its anti-brane which are stretched in $X$.  The charged branes are
charged under the single gauge field of the IIA theory.

\bigskip
\centerline{\bf Acknowledgements}

We thank M.~Fabinger, N.~Itzhaki, D.~Kutasov, J.~Maldacena,
S.~Murthy, Y.~Oz, D.~Shih, E.~Silverstein and E.~Witten for
discussions. This work was supported in part by grant
\#DE-FG02-90ER40542.

\listrefs
\end